\documentclass{eas}

\usepackage{graphicx}
\usepackage{subfig}
\usepackage{natbib}

\begin{document}

\title{Least-Squares Deconvolution based analysis of stellar spectra}
\runningtitle{Van Reeth \etal: LSD-based analysis of stellar spectra}

\thanks{We are also grateful to Conny Aerts (KU Leuven / University of Nijmegen), Oleg Kochukhov (Uppsala University) and Jonas Debosscher (KU Leuven), for the very helpful discussions.}

\author{T. Van Reeth}\address{Instituut voor Sterrenkunde, KU Leuven, Celestijnenlaan 200D, B-3001 Leuven, Belgium}
\author{A. Tkachenko}\sameaddress{1}
\author{V. Tsymbal}\address{Tavrian National University, Department of Astronomy, Simferopol, Ukraine}

\begin{abstract}
 In recent years, astronomical photometry has been revolutionised by space missions such as MOST, CoRoT and \emph{Kepler}. However, despite this progress, high-quality spectroscopy is still required as well. Unfortunately, high-resolution spectra can only be obtained using ground-based telescopes, and since many interesting targets are rather faint, the spectra often have a relatively low S/N. Consequently, we have developed an algorithm based on the least-squares deconvolution profile, which allows to reconstruct an observed spectrum, but with a higher S/N. We have successfully tested the method using both synthetic and observed data, and in combination with several common spectroscopic applications, such as e.g. the determination of atmospheric parameter values, and frequency analysis and mode identification of stellar pulsations.
\end{abstract}

\maketitle
\section{Introduction}
Thanks to space missions such as MOST, CoRoT and \emph{Kepler},
there have been many advances in the field of photometry over the
last decade. However, astronomical research still fundamentally
depends on spectroscopy as well. On one hand, it allows us to, e.g.,
obtain values for the fundamental stellar parameters, while on the
other, time series of spectra with a relatively high S/N often allow
us to analyse stellar variability or binarity. It is also for this
reason that, in the case of double-lined spectroscopic binaries,
disentangling composite spectra is such a valuable technique: it
allows us to gain detailed information about the individual
components, which is nigh-on impossible for photometric data.
Unfortunately, we currently need ground-based observations to obtain
high-resolution spectra, and since many of the interesting objects
are faint, ofttimes 4m-class telescopes are required to properly
observe targets with a high S/N. However, as observation time on
these instruments is limited, observers often have to make due with
2m-class telescopes, resulting in lower S/N data.

With this shortcoming in mind, we developed a new algorithm, based
on the least-squares deconvolution (LSD) profile, with the aim to
filter out noise of observed high-resolution spectra. In these
proceedings we only provide a concise description of the algorithm
and several of the conducted tests, and we refer the interested
reader to \citet{Tkachenko2013} for more detailed information.

\section{The Least-Squares Deconvolution algorithm}
\subsection{The classical LSD profile}
\label{subsec:LSD}

The basic idea of the LSD technique is to extract an ``average''
high-precision line profile from stellar intensity or polarisation
spectra. The method was first proposed and developed by
\citet{Donati1997}, and is based on two assumptions: (i) individual
spectral lines have the same intrinsic line profile, and (ii)
overlapping spectral lines add up linearly. We can therefore write
the spectrum $Y$ as $$Y\ \ =\ \ M*Z(v),$$ where $M = \sum_i
w_i\delta(v-v_i)$ is the line pattern function (with $w_{i}$ the
central line depths of the individual lines) and $Z(v)$ is the
intrinsic line profile. As described by \citet{Donati1997} and
\citet{Kochukhov2010}, this equation can be rewritten using matrices
$$\mathbf{Y}\ \ =\ \ \mathbf{M}\cdot\mathbf{Z},$$ which in turn can
be expressed as the inverse problem $$\mathbf{Z}\ \ =\ \
\left(\mathbf{M}^T\cdot\mathbf{S}^2\cdot\mathbf{M}\right)^{-1}\cdot\left(\mathbf{M}^T\cdot\mathbf{S}^2\cdot\mathbf{Y}_
0\right).$$ Here $\mathbf{Y}_0$ is the observed spectrum and
$\mathbf{S}^2$ is a diagonal weighting matrix based on the variance
of the data points in the spectrum. We note that, by solving for the
LSD profile $\mathbf{Z}$, we effectively deconvolve the weighted
cross-correlation vector with the auto-correlation matrix. This
allows us to remove the spurious signal which might be present in
the cross-correlation due to regularities in the used line pattern
function.

Unfortunately, the multiplication with the numerically computed
inverse of the matrix also tends to amplify the noise which is
present in the cross-correlation. To avoid this complication, we
chose to solve the inverse problem using the Levenberg-Marquardt
algorithm. We do not make any assumptions beforehand on either the
shape or the depth of the LSD-profile, but take the continuum flux
level at unity as the initial guess for the profile. And in order to
limit the computation time, we use a fast modified version of the
Levenberg-Marquardt algorithm (developed by \citealt{Piskunov2002}).

\subsection{Generalisation of the technique}
\subsubsection{The multi-profile technique}

Aside from the numerical calculations, the LSD algorithm itself can
also be improved. The intrinsic line shape, for instance, actually
varies from line to line, e.g. dependent on the pressure broadening.
To take this into account, \citet{Kochukhov2010} assumed each
spectral line to be a linear combination of different intrinsic line
shapes, each of which corresponds to a different line pattern
function and can be located in different parts of the velocity
space:
\begin{equation}Y\ \ =\ \ \sum_jM_j*Z_j(v).\label{eq:specrec}\end{equation}

\begin{figure}
 \centering
 \includegraphics[width=0.57\textwidth]{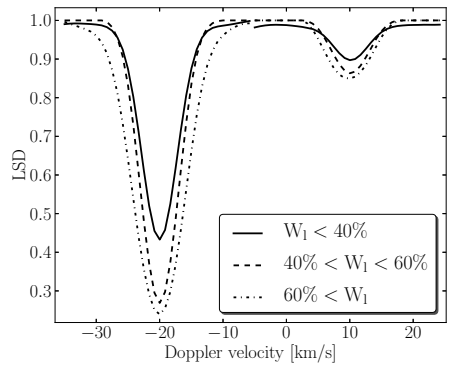}
 \caption{\label{fig:multip}The LSD profiles computed for a synthetic composite spectrum of a binary ($L_1/L_{\rm tot} = 85\%$; primary: $T_{\rm eff} = 8\,500$~K, $\log g = 3.4$~dex, [M/H] = 0.0~dex, $v \sin i = 2$~km\,s$^{-1}$; secondary: $T_{\rm eff} = 7\,300$~K, $\log g = 3.5$~dex, [M/H] = -0.3~dex, $v \sin i = 2$~km\,s$^{-1}$). The line strength $w_{l}$ is expressed here as a percentage of the continuum flux.}

\end{figure}

This allows us, for instance, to properly apply the LSD algorithm on
composite spectra of an SB2 consisting of components with different
atmospheric parameter values. And by computing different LSD
profiles for the spectral lines of a single star, in function of the
line depths, we can also partially take into account variations in
the pressure broadening of spectral lines. (See Figure
\ref{fig:multip}.) Unfortunately, the multi-profile implementation
of the LSD technique is still insufficient to model hydrogen,
helium, and strong metal lines with dampened wings. As a
consequence, our generalised LSD technique is only applicable for
stellar spectra with a large number of usable metal lines. So far
the algorithm has been tested for stars with a $T_{\rm eff}$ between
5\,000~K and 10\,000~K. The performance of the code for other stars
is unknown.

\subsubsection{Line strength correction}

\begin{figure}
 \centering
 \includegraphics[width=0.57\textwidth]{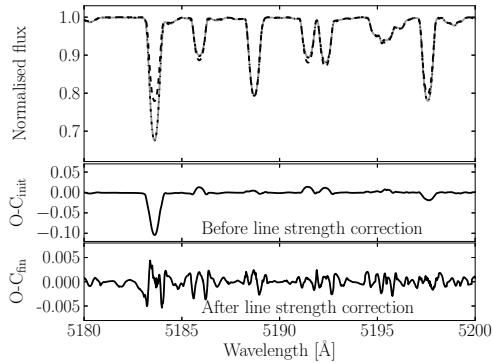}
 \caption{\label{fig:lincor}The spectral reconstruction for a synthetic spectrum ($T_{\rm eff} = 8\,500$~K, $\log g = 3.4$~dex, [M/H] = 0.0~dex, $\xi = 2$~km\,s$^{-1}$ and $v\sin i = 20$~km\,s$^{-1}$). The upper plot shows the input spectrum (grey full line), and the LSD reconstructions before (dashed) and after (dash-dotted) the line strength adaptation. Note also the scaling of the axes for the O-C data.}
\end{figure}

If we now compute the convolution of the obtained LSD profiles and
the lines in the line pattern functions, we should be able to
reconstruct the observed spectrum, but with a much higher S/N (as we
can see from equation \ref{eq:specrec}). However, many spectral
lines in the observed spectrum are likely to have central line
depths which differ from the values used in the line pattern
function. In addition, we incorrectly assumed that blended lines add
up linearly (see Section \ref{subsec:LSD}): e.g. line blending due
to pressure broadening of the line profiles is non-linear. As a
consequence, there will be discrepancies between the observed and
reconstructed spectrum. In order to compensate for these
shortcomings, we ``optimise'' the spectral line strengths by fitting
the reconstructed spectrum to the observed one. (See Figure
\ref{fig:lincor}). The strengths of the spectral lines are adapted
one at a time (by means of the golden section search algorithm),
taking into account the contributions from the other lines. If,
however, neighbouring spectral lines cannot be resolved by the used
spectrograph, we do not treat them separately, but solve for the sum
of their line strengths instead.

To allow all of the line strengths to properly converge towards the
solution, we iterate several times over the complete line list
(typically ten times). Afterwards, we recompute the LSD profiles,
and iterate globally over all the computations, until neither the
LSD profiles nor the line pattern functions vary significantly
anymore. Typically, two global iterations of the entire computation
cycle is sufficient.

\section{Applications}

While the original goal of this research was mainly to analyse to
binaries, we first tested the algorithm on both synthetic and
observed spectra of single stars, before taking into account the
effect of binarity. Both the synthetic spectra and the line masks
(i.e. the line pattern functions) used in these tests were computed
with the {\sc synthv} code \citep{Tsymbal1996}. The atmosphere
models we used in the {\sc synthv} code were computed with the {\sc
llmodels} code \citep{Shulyak2004}, based on information extracted
from the {\sc vald} database \citep{Kupka1999}.
\subsection{Single stars}
\subsubsection{Atmospheric parameter value determination}
\label{subsubsec:atmos}

The first application test we conducted was the determination of
atmospheric parameter values, for both synthetic and observational
data. In both cases, the parameter value computations were done
using the {\sc gssp} code \citep{Lehmann2011}. For the run with the
synthetic spectra we took $T_{\rm eff} = 8\,500$~K, $\log g =
3.4$~dex, [M/H] = 0.0~dex, $\xi = 2$~km\,s$^{-1}$ and $v\sin i =
2$~km\,s$^{-1}$ and added varying degrees of noise (S/N = 35; 70;
$\infty$), while the atmospheric parameters of the line masks varied
near those of the input spectra ($T_{\rm eff} \pm 500$~K, $\log g
\pm 0.5$~dex, [M/H]$\pm$0.3~dex). By comparing the atmospheric
parameter values computed from the LSD reconstructions with the ones
from the synthetic input spectra, we were able to also study the
precise influence of the chosen line pattern function on the end
result. The test results then showed that the differences between
the LSD reconstructed parameter values and those of the input
spectrum were systematically smaller than the error margins
(obtained using $\chi^2$~statistics). The largest deviations were
obtained when the metallicity of the used mask differed from the
metallicity of the input spectrum.

\begin{figure}
 \centering
\subfloat[b][for Vega]{\begin{minipage}[t]{0.48\textwidth}
                 \label{fig:Vega}\includegraphics[width=\textwidth]{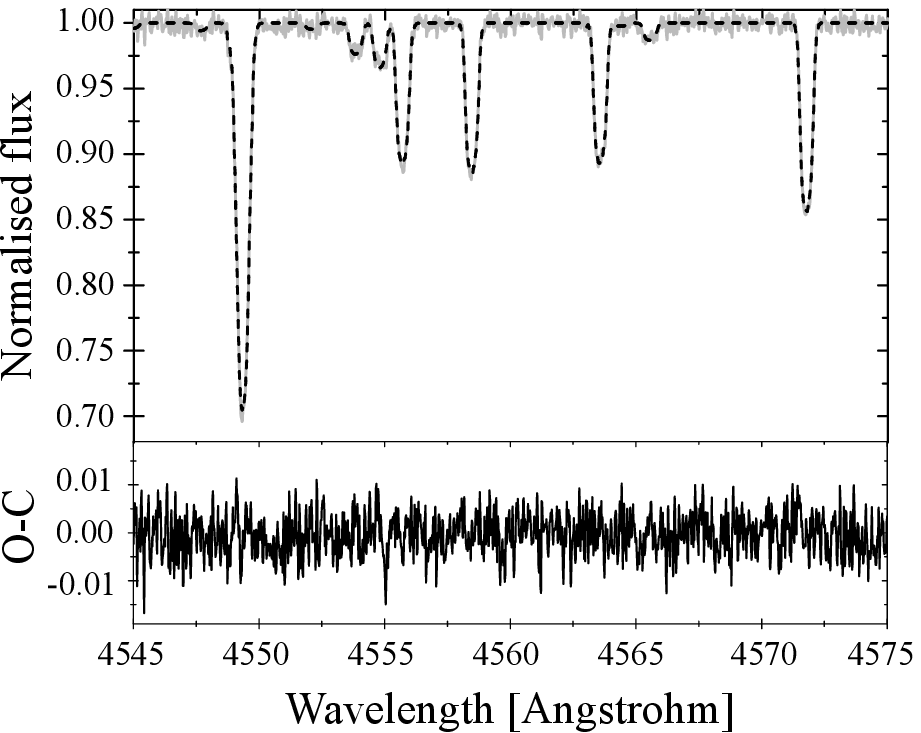}
                \end{minipage}}
\hspace{0.02\textwidth}
\subfloat[b][for KIC~4749989]{\begin{minipage}[t]{0.48\textwidth}
                        \vspace{5.25mm}\label{fig:KIC4749989}\includegraphics[width=\textwidth]{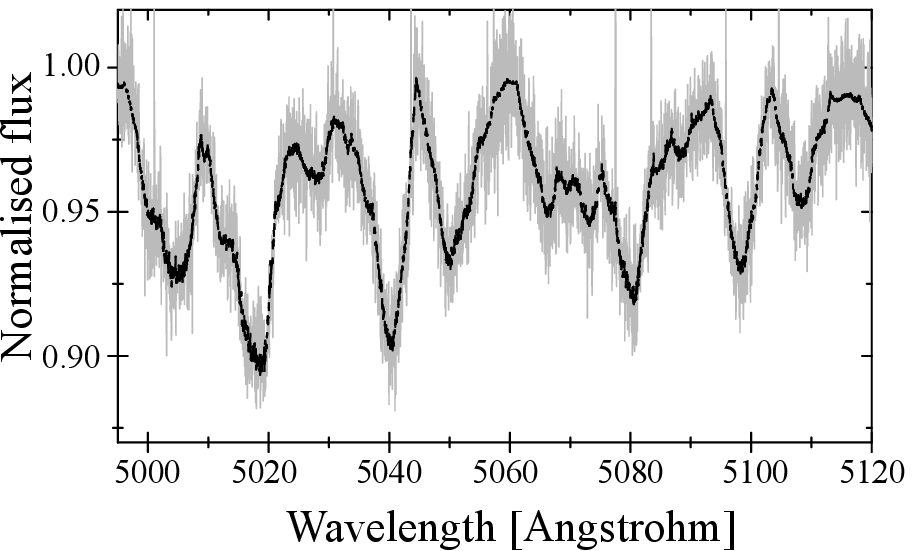}\\
                        \vspace{2mm}
                       \end{minipage}}
\caption{\label{fig:obsspec}A comparison of the observed (grey) and LSD reconstructed (black) spectra}
\end{figure}

The application of the LSD algorithm on observational data provided
similar results. We used spectra of Vega ($v\sin i =
22$~km\,s$^{-1}$, S/N $\sim$ 100) and KIC\,4749989 ($v\sin
i~=~190$~km\,s$^{-1}$, S/N $\sim$ 60), taken with the {\sc hermes}
spectrograph at the 1.2m Mercator Telescope (Observatorio del Roque
de los Muchachos, La Palma). As illustrated in Figure
\ref{fig:obsspec}, the observed spectra were reconstructed very
well, and the atmospheric parameter values computed from these LSD
reconstructions agreed with the values found in the literature
within the 1$\sigma$ confidence level.

\subsubsection{Intrinsically variable stars}

Next, we applied the LSD algorithm on stellar spectra showing
intrinsic line profile variations due to pulsations. The first test
case was 20~CVn, a mono-periodic radial $\delta$ Scuti pulsator
\citep[e.g.][]{Rodriguez1998}. The spectra were retrieved from the
archive of the high-resolution \'echelle spectrograph {\sc elodie}
($R$ = 42\,000, 67 orders, 3895-6815\AA) at the 193cm-telescope at
the Observatoire de Haute-Provence (OHP), which was in use from 1993
to 2006. We worked with the ScII-line at 5239\AA, and by adding
gaussian white noise to the observed data (S/N $\sim 40$) before
applying the LSD algorithm, we saw that, while the quality of the
noisy ``original'' data was insufficient to do either a frequency
analysis or pulsational mode identification, the time series of LSD
reconstructed ScII-lines showed no such limitations. For our other
test case, HD\,189631, a multi-periodic non-radial $\gamma$ Doradus
pulsator, we obtained equally successful results as we did for
20~CVn.

These tests showed that, if the spectral lines show the same
response to the pulsations and assuming we have enough observations
to resolve the oscillation frequencies, we can determine the
oscillation frequency values and do a mode
identification\footnote{Mode identification results obtained with
the software package FAMIAS developed in the framework of the FP6
European Coordination Action HELAS. (http://www.eu-helas.org/)} from
either the LSD profiles themselves, or from a (properly)
reconstructed non-blended spectral line \citep{Zima2008}.

\subsection{Binary stars}

Thanks to the multi-profile implementation of the LSD technique, we
are also able to properly take into account binarity, which allows
us to facilitate the analysis of noisy composite spectra of SB2s. A
good example is KIC\,11285625, a double-lined binary where the
primary shows $\gamma$ Dor-type pulsations \citep{Debosscher2013}.
In the original paper, the authors obtained the disentangled spectra
of both binary components by means of the {\sc fdbinary} code
\citep{Ilijic2004}, but the spectrum of the secondary was of too low
S/N the atmospheric parameters determination to be possible.
\citet{Tkachenko2013} showed that application of our LSD technique
to the original composite spectra and application of the spectral
disentangling to the LSD model spectra afterwards leads to
significant increase in S/N of both decomposed spectra providing a
good enough quality of the data to make detailed spectrum analysis
of both binary components possible.

\subsubsection{Spectral disentangling}
The analysis by \citet{Debosscher2013} again confirms the usefulness
of spectral disentangling. It is therefore also worth noting that it
is possible to use the LSD algorithm to do spectral disentangling:
we only have to make sure that, when we convolve the LSD profiles
and the line pattern functions in equation \ref{eq:specrec}, we do
this for each contributing stellar component individually.
Unfortunately however, the spectral line strength correction which
is applied afterwards, also introduces a degeneracy in the solution,
since most of the spectral lines of the primary and the secondary
overlap. As a result, proper spectral disentangling based on the LSD
algorithm is currently not yet possible, though we have several
ideas on how to solve this problem.

\subsubsection{Light factor estimation}

Applying the LSD algorithm to the composite spectra of SB2s, can
further provide us with an estimate of the light factor $\alpha$ of
the system, if we assume that $$\alpha\ \ =\ \
\frac{EW^{LSD}_1}{EW^{LSD}_1 + EW^{LSD}_2}.$$ We tested this
hypothesis by taking a synthetic composite binary spectrum with the
same atmospheric parameters as for the LSD profiles in Figure
\ref{fig:multip} (primary: $T_{\rm eff} = 8500$~K, $\log g =
3.4$~dex, [M/H] = 0.0~dex; secondary: $T_{\rm eff} = 7300$~K, $\log
g = 3.5$~dex, [M/H] = -0.3~dex), and then allowing the atmospheric
parameter values of the used line masks to vary (similar to the
tests we conducted in Section \ref{subsubsec:atmos}) to test the
influence of the used line masks.

In most cases the light factor $\alpha$ could be recovered with an
accuracy of about 5\% of the continuum. Though there is an exception
when the metallicities of the used line masks differs from the
metallicities of the input spectrum: then the error of the recovered
light factor can go up to about 12\%.

\section{Conclusions \& future work}

With the launches of high-performance space missions, astronomical
research based on photometry has been revolutionised. However, we
still require spectroscopic observations as well, since a lot of
useful information, such as e.g. fundamental stellar parameter
values cannot be determined from the analysis of white-light
photometric light curves. Unfortunately, many interesting targets
are faint, and due to a limited observation time, ofttimes 2m-class
telescopes have to be used, resulting in rather noisy spectra. We
have developed an algorithm, based on the LSD technique, which
allows us to reconstruct stellar spectra of both single and binary
stars, but with a much higher S/N. So far the method has been tested
for stars with $5\,000~{\rm K} \leq T_{{\rm eff}} \leq 10\,000~{\rm
K}$, and has been shown to be compatible with several other spectral
analysis techniques, including the determination of atmospheric
parameter values, the analysis of line profile variations, and in
the case of binaries, Fourier spectral disentangling and estimating
the light factor $\alpha$. However, as we have seen, there is also
still a lot of work to be done. When applying the LSD algorithm
itself to do spectral disentangling, we currently suffer from
degeneracy. And at this moment, we do not know how the algorithm
performs for stars with a temperature $T_{\rm eff}$ lower than
$5\,000$~K or higher than $10\,000$~K. It is clear that further
testing of the LSD method is paramount, and anyone interested in
collaborating on the subject is welcome to contact us.

\bibliographystyle{astron}

\begin{thebibliography}{}

\bibitem[\protect\astroncite{{Debosscher} \etal\/}{2013}]{Debosscher2013}
{Debosscher}, J., {Aerts}, C., {Tkachenko}, A. \etal\/ 2013, A\&A, 556, A56

\bibitem[\protect\astroncite{{Donati} \etal\/}{1997}]{Donati1997}
{Donati}, J.-F., {Semel}, M., {Carter}, B.~D. \etal\/ 1997, MNRAS, 291, 658

\bibitem[\protect\astroncite{{Iliji\'{c} \etal\/}}{2004}]{Ilijic2004}
{Iliji\'{c}}, S., {Hensberge}, H., {Pavlovski}, K., \& Freyhammer, L.~M. 2004, ASPC, 318, 111

\bibitem[\protect\astroncite{{Kochukhov} \etal\/}{2010}]{Kochukhov2010}
{Kochukhov}, O., {Makaganiuk}, V., \& {Piskunov}, N. 2010, A\&A, 524, A5

\bibitem[\protect\astroncite{{Kupka} etal\/}{1999}]{Kupka1999}
{Kupka}, F., {Piskunov}, N., {Ryabchikova}, T.~A. \etal\/ 1999, A\&AS, 138, 119

\bibitem[\protect\astroncite{{Lehmann} \etal\/}{2011}]{Lehmann2011}
{Lehmann}, H., {Tkachenko}, A., {Semaan}, T. \etal\/ 2011, A\&A, 526, A124

\bibitem[\protect\astroncite{{Piskunov} \& {Kochukhov}}{2002}]{Piskunov2002}
{Piskunov}, N., \& {Kochukhov}, O. 2002, A\&A, 381, 736

\bibitem[\protect\astroncite{{Rodr\'{\i}guez} \etal\/}{1998}]{Rodriguez1998}
{Rodr\'{\i}guez}, E., {Rolland}, A., {Garrido}, R. \etal\/ 1998, A\&A, 331, 171

\bibitem[\protect\astroncite{{Shulyak} \etal\/}{2004}]{Shulyak2004}
{Shulyak}, D., {Tsymbal}, V., {Ryabchikova}, T. \etal\/ 2004, A\&A, 428, 993

\bibitem[\protect\astroncite{Tkachenko \etal\/}{2013}]{Tkachenko2013}
Tkachenko, A., {Van Reeth}, T., Tsymbal, V. \etal\/ 2013, A\&A, accepted (arXiv1310.3198T)

\bibitem[\protect\astroncite{{Tsymbal}}{1996}]{Tsymbal1996}
{Tsymbal}, V. 1996, ASPC, 108, 198

\bibitem[\protect\astroncite{{Zima}}{2008}]{Zima2008}
{Zima}, W. 2008, CoAst, 155, 17

\end{thebibliography}

\end{document}